\documentclass[acmsmall]{acmart}

\usepackage{booktabs}
\usepackage{array}
\usepackage{makecell}
\usepackage{xcolor}
\usepackage{graphicx}
\usepackage{listings}
\usepackage{tikz}
\usepackage{pgfplots}
\pgfplotsset{compat=1.16}
\usetikzlibrary{arrows.meta,positioning,fit,calc,shapes.geometric}

\acmJournal{TODAES}
\graphicspath{{figs/}}

\newif\ifanon\anonfalse

\newcommand{\framework}{SABLE}
\newcommand{\bridge}{SafeBridge}
\newcommand{\pass}{\textsc{Pass}}
\newcommand{\fail}{\textsc{Fail}}
\newcommand{\metric}[1]{\texttt{#1}}
\newcommand{\cmark}{\checkmark}
\newcommand{\xmark}{\ensuremath{\times}}

\lstdefinestyle{jsonstyle}{
  basicstyle=\ttfamily\footnotesize,
  breaklines=true,
  frame=single,
  framerule=0.3pt,
  columns=fullflexible,
  keepspaces=true,
}

\title{\framework: An NDA-Safe Closed-Loop LLM Framework for Analog Circuit Optimization in Industrial EDA Flows}

\author{Xunqi Li}
\affiliation{%
  \institution{University of Minnesota, Twin Cities}
  \city{Minneapolis}\country{US}}
\email{li001674@umn.edu}

\author{Chris H. Kim}
\affiliation{%
  \institution{University of Minnesota, Twin Cities}
  \city{Minneapolis}\country{US}}
\email{chriskim@umn.edu}

\begin{document}

\begin{abstract}
Large language models (LLMs) can propose circuit-sizing decisions, but
industrial analog design flows cannot simply expose foundry process-design-kit
(PDK) content, proprietary schematic details, absolute simulation paths, or
license-bound tool state to a cloud endpoint. This paper presents
\framework{} (Safe Analog Boundary for LLM-driven EDA), an NDA-safe
closed-loop framework that lets LLMs optimize
analog circuits through Cadence Virtuoso, Maestro, and Spectre while
returning only scrubbed topology intent, numeric simulation metrics,
operating-point summaries, and scoped writeback status. We use
``NDA-safe'' to mean enforcement under a stated curious-but-passive
cloud-provider threat model, not a formal non-interference proof. The framework
combines an explicit threat model for the cloud-LLM boundary, a
whitelist of 28 scoped SKILL entry points, PDK/path/model scrubbing on
every return path, structured Maestro setup and writeback recipes, a
strict JSON action contract with six machine-checked stop conditions,
and best-so-far state preservation. We evaluate eleven LLM checkpoints
from the same documented reset state on two real closed-loop tasks, both run as
process--voltage--temperature (PVT) sign-offs across the same three corners: a
20\,GHz LC-VCO tuning-curve optimization task and a differential two-stage op-amp
optimization task. On the LC-VCO task, 7 of 11 models pass; on the harder op-amp
task---where every metric must hold at the worst of three corners and a
phase-margin gate rejects unstable high-gain points---4 of 11 models pass within
the 15-iteration budget. Single-checkpoint feedback-path ablations show that
removing individual sanitized feedback channels either silently weakens the
specification or measurably degrades the search. The results show that model quality differs sharply once the
loop requires tool discipline, bias reasoning, and specification repair,
and that an NDA-safe EDA boundary can still provide enough sanitized
feedback for successful analog circuit optimization.
\end{abstract}



\maketitle

\begin{figure}[t]
  \centering
  \includegraphics[width=\linewidth]{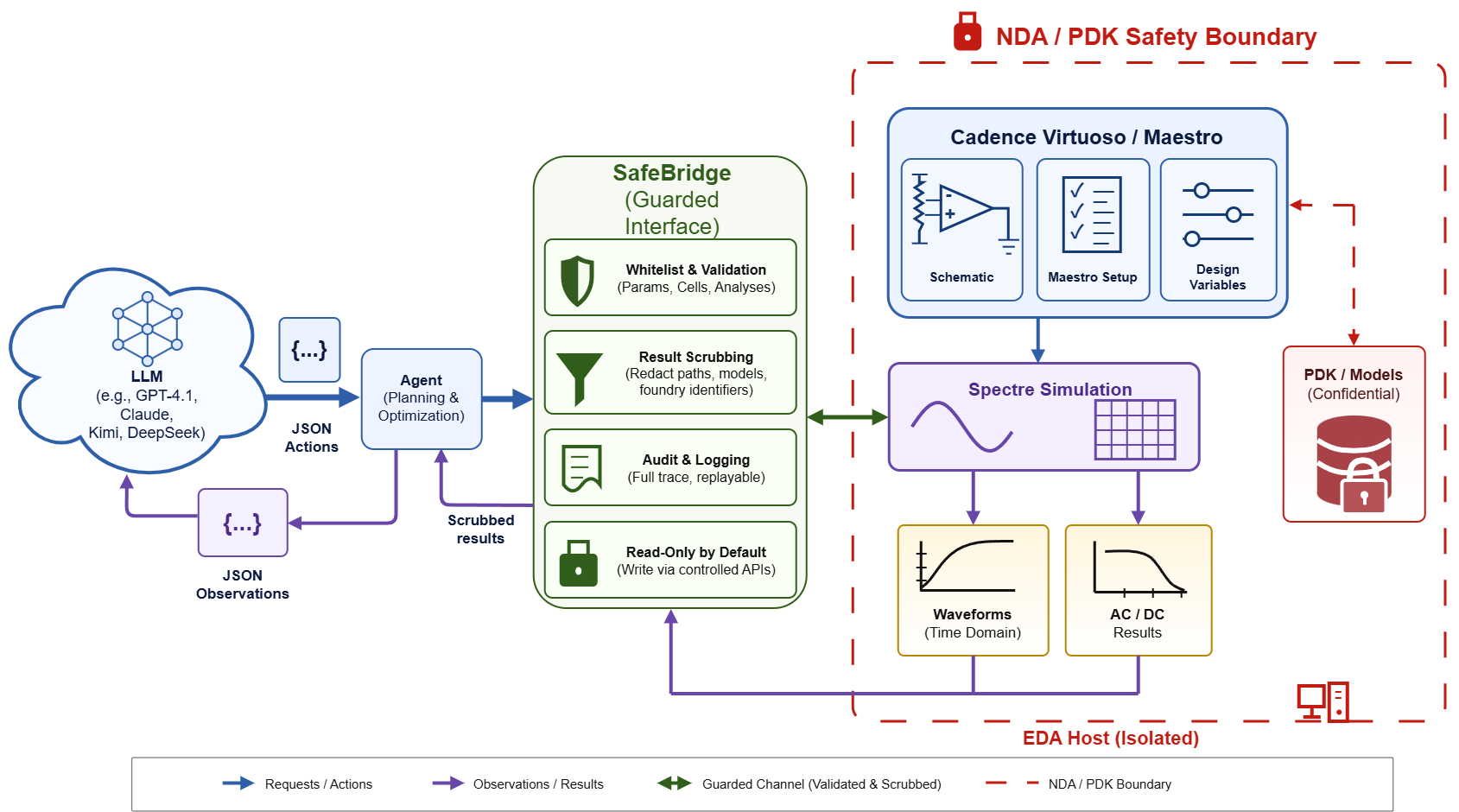}
  \caption{Conceptual overview of \framework. The LLM sees JSON actions and
  scrubbed observations. Foundry PDK content, Maestro state, simulator raw
  paths, and proprietary model details remain inside the isolated EDA host.}
  \Description{Conceptual diagram. A cloud LLM exchanges JSON actions and
  scrubbed observations with the Python agent across a trust boundary; the
  agent drives an isolated EDA host running Cadence Virtuoso, Maestro, and
  Spectre, while foundry PDK content, raw simulation paths, and model details
  stay inside the host.}
  \label{fig:teaser}
\end{figure}

\section{Introduction}
\label{sec:intro}

LLM-driven EDA loops are useful only when the tool boundary is as carefully
engineered as the prompt. Recent digital, HLS, and full-flow EDA agents use
compiler, synthesis, place-and-route, or simulation feedback to refine
LLM-generated artifacts \cite{autochip,hlsrewriter,c2hlsc,lhs,confibench,
chateda}. Analog circuit optimization is a natural next target because SPICE
simulation provides a direct numerical oracle. However, the analog case
introduces a deployment constraint that is not a minor implementation detail:
a real foundry-bound Virtuoso flow contains PDK model cards, proprietary cell
names, private schematic hierarchy, absolute host paths, and license-bound
simulation state that cannot be sent to a cloud LLM.

Existing analog LLM work has made rapid progress on code generation, topology
synthesis, sizing, testbench construction, and benchmark design
\cite{analogcoder,analogcoderpro,circuitsynth,analogxpert,adollm,eesizer,
anaflow,ledro,autosizer,analogsage,analoggym,aicircuit,amsnet,masalachai}. The
missing systems question is how to let a frontier model use the feedback it
needs without crossing an NDA boundary. Open-PDK benchmarks, public netlist
datasets, domain-adapted chip models, and on-premise model deployments are
valuable, but they leave two practical questions unanswered for industrial
users: can cloud LLMs be connected to a real Virtuoso/Maestro loop without
leaking PDK content, and how should models be compared when every run must
start from the same documented EDA state?

This paper argues that the right unit of contribution is not another
standalone prompt or a claim that an LLM is a circuit designer. The unit is a
reproducible, safety-bounded analog circuit-optimization substrate. \framework{} wraps the
Cadence SKILL and OCEAN interface with a Python agent, a return-path scrubber,
per-project design-variable whitelists, structured Maestro setup recipes,
simulation-result readback, and a strict JSON action contract
(Figure~\ref{fig:teaser}). The LLM can
propose numeric design variables and reason over sanitized topology and
operating-point summaries, but it cannot browse arbitrary libraries, read raw
PDK model files, or receive unsanitized simulator paths. Crucially, the
boundary is not an informal convention: it is specified as a threat model
with named protected assets, an enumerated attack ladder, and five invariants
that are enforced on every bridge call and exercised by red-team unit tests.

We make five contributions:
\begin{enumerate}
  \item We introduce an NDA-safe closed-loop architecture for LLM-driven analog
  circuit optimization in Virtuoso/Maestro flows, together with an explicit threat model: a
  curious-but-passive cloud adversary, a four-tier attack ladder with the
  enforcement status of each tier, and five hard invariants checked on every
  call (Sections~\ref{sec:threat} and \ref{sec:framework}).
  \item We define a machine-checkable interaction protocol: a strict JSON
  action contract with an engineering-suffix value grammar, six enumerated
  stop conditions, scoped Maestro writeback with landing verification, and
  best-so-far state preservation (Section~\ref{sec:framework}).
  \item We define a fair benchmark protocol that resets each model to the same
  documented initial design-variable state, runs the same Maestro tests, and
  logs the terminal cause of every run (Section~\ref{sec:protocol}).
  \item We report an eleven-model matched comparison on two circuit classes,
  both run as process--voltage--temperature (PVT) sign-offs with worst-corner
  gating across the same three corners---an LC-VCO tuning-curve task and a
  two-stage op-amp optimization task---with full per-run artifacts
  (Section~\ref{sec:results}).
  \item We provide initial feedback-path ablation evidence and a failure
  taxonomy with observed terminal-cause counts, showing where closed-loop
  analog agents break: contract violations, missing tool discipline, wrong
  bias-scale reasoning, insufficient use of operating-point evidence, and
  endpoint or harness failures (Sections~\ref{sec:ablation}
  and \ref{sec:failures}).
\end{enumerate}

The rest of the paper is organized as follows. Section~\ref{sec:related}
positions the work. Section~\ref{sec:threat} defines the threat model.
Section~\ref{sec:framework} describes the closed-loop framework and the
interaction protocol. Section~\ref{sec:protocol} defines the benchmark tasks
and the fair-start protocol. Section~\ref{sec:results} reports the
eleven-model comparison. Section~\ref{sec:ablation} reports feedback-path
ablation evidence. Section~\ref{sec:failures} analyzes failures.
Section~\ref{sec:repro} describes the reproducibility package and security
audit, and Sections~\ref{sec:discussion} and \ref{sec:conclusion} conclude.

\section{Related Work}
\label{sec:related}

\subsection{LLMs in EDA}

LLM4EDA systems have been most visible in digital RTL, HLS, and code-oriented
flows, where the feedback loop can be closed with compiler and synthesis
diagnostics. AutoChip, HLSRewriter, C2HLSC, LHS, and ConfiBench demonstrate
this pattern \cite{autochip,hlsrewriter,c2hlsc,lhs,
confibench}, while ChatEDA broadens the same agentic idea toward an RTL-to-GDS
tool-execution flow \cite{chateda}. Domain adaptation work such as ChipNeMo
also shows that chip-design-specific context can improve LLM usefulness across
assistant, script-generation, and debugging tasks \cite{chipnemo}. These works
motivate the central premise of this paper: LLMs become more useful when the
EDA tool is in the loop and the feedback is machine-checkable. What the
digital results do not transfer, however, is the confidentiality posture: a
Verilog compile log can usually be shared with a cloud endpoint, whereas a
Spectre run on a foundry PDK cannot.

\subsection{LLMs for analog design}

Analog LLM work has moved from netlist/code generation toward sizing and
workflow automation. AnalogCoder and AnalogCoder-Pro focus on analog circuit
generation and optimization \cite{analogcoder,analogcoderpro}. ADO-LLM,
EEsizer, AnaFlow, LEDRO, AutoSizer, and AnalogSAGE explore LLM-aided sizing,
multi-agent search, memory, and optimization strategies
\cite{adollm,eesizer,anaflow,ledro,autosizer,analogsage}. CircuitSynth and
AnalogXpert study LLM-based topology synthesis, while Circuit-Agent and
AnalogAgent represent broader agentic analog-design workflows
\cite{circuitsynth,analogxpert,circuitagent,analogagent}. AnalogGym, AICircuit,
AMSNet, AMSnet-KG, and Masala-CHAI expand the benchmark and dataset side of
the field \cite{analoggym,aicircuit,amsnet,amsnetkg,masalachai}. These efforts
establish that LLMs can participate in analog design loops. They largely
operate on open PDKs, exported netlists, or simulator wrappers that assume
the model may see the design database. Our contribution is complementary: we
target the safety and reproducibility boundary required by foundry-bound
Virtuoso/Maestro deployment, where the schematic, the PDK, and the simulator
host must remain opaque to the model. In particular, AMSnet-KG
\cite{amsnetkg}, a concurrent effort, performs knowledge-graph-retrieval
netlist auto-design on a foundry (28\,nm) PDK with Cadence Spectre and
Bayesian-optimization sizing---the closest prior setting to ours---yet it
defines no model-provider trust boundary, so it is orthogonal to the scrubbed
cloud-LLM deployment substrate studied here.

Table~\ref{tab:related} positions \framework{} against representative
analog-design systems along the axes that matter for a foundry-bound, cloud-LLM
deployment. Several prior systems already iterate over simulation feedback,
expose operating-point hints, or compare multiple models; one even runs a real
foundry PDK through Cadence Spectre. What none of them do is place that licensed
Cadence flow behind an \emph{enforced model-provider trust boundary}. That
boundary---not another optimizer---is our contribution.

\begin{table}[t]
  \centering
  \caption{Where \framework{} sits relative to representative analog-design
  systems. \emph{Real Cad.}: drives a licensed Cadence (Virtuoso/Maestro/Spectre)
  flow rather than open-source or wrapped simulators; \emph{Safe bnd.}: states
  and enforces a model-provider trust boundary with return-path scrubbing;
  \emph{Wrt.bk.}: iteratively writes sized variables back and re-simulates;
  \emph{OP rdbk.}: exposes sanitized per-device operating-point evidence;
  \emph{Multi-mdl.}: matched benchmark across multiple LLMs. Entries reflect a
  direct reading of each cited paper.}
  \label{tab:related}
  \scriptsize
  \setlength{\tabcolsep}{3pt}
  \begin{tabular}{l c c c c c c}
    \toprule
    & \makecell{PDK} & \makecell{Real\\Cad.} & \makecell{Safe\\bnd.} &
    \makecell{Wrt.\\bk.} & \makecell{OP\\rdbk.} & \makecell{Multi\\mdl.} \\
    \midrule
    AnalogCoder \cite{analogcoder}         & Open  & \xmark & \xmark & \cmark & \cmark & \cmark \\
    AnalogAgent \cite{analogagent}         & ---   & \xmark & \xmark & \cmark & \cmark & \cmark \\
    AMSnet-KG \cite{amsnetkg}              & Priv. & \cmark & \xmark & \cmark & \xmark & \xmark \\
    AnalogGym$^{\dagger}$ \cite{analoggym} & Both  & \xmark & \xmark & \cmark & \xmark & \xmark \\
    \midrule
    \framework{} (ours)                    & Priv. & \cmark & \cmark & \cmark & \cmark & \cmark \\
    \bottomrule
  \end{tabular}

  \smallskip
  \raggedright\scriptsize $^{\dagger}$AnalogGym is a sizing-\emph{algorithm}
  benchmark (BO/RL/EA), not an LLM agent; listed as the closest open evaluation
  suite. AMSnet-KG's \emph{Real Cad.} is batch Spectre, without the
  Virtuoso/Maestro ADE flow that \framework{} targets.
\end{table}

\subsection{NDA-constrained deployment}

Two broad strategies can address NDA-constrained analog LLM use. One is to
move more model capability inside the trusted environment through
domain-adapted, local, or self-improving agents \cite{chipnemo,analogagent}.
The other is to keep cloud models outside the trusted environment and strictly
control what crosses the boundary. \framework{} takes the second path. It does
not claim to remove the need for local EDA licenses or foundry access; instead,
it defines a scrubbed bridge that lets external models see enough evidence to
optimize without seeing the confidential artifacts that produced the evidence.
To our knowledge, no prior analog LLM sizing work states an explicit threat
model for the model-provider boundary or audits its feedback path against one.

\section{Threat Model and Safety Boundary}
\label{sec:threat}

\subsection{Adversary and protected assets}

The framework assumes that the cloud LLM provider is
\emph{curious-but-passive}: it can observe every prompt and returned tool
result and may log, train on, or later mine that traffic offline, and the tool
arguments it returns may be adversarial; but it does not compromise the local
operating system or the Cadence process, and it does not mount active
man-in-the-middle attacks on the EDA host. ``Passive'' here refers to the
absence of in-band host compromise, not to benign intent---adversarial tool
arguments are squarely in scope and are handled by the Tier-2 defenses below. Under this model, the protected assets are:
\begin{itemize}
  \item foundry device-model content (model-card parameter names and values,
  e.g., BSIM-class threshold, mobility, and capacitance parameters);
  \item process back-end information (layer stack, pitches, parasitics);
  \item proprietary cell and library identifiers beyond the allowed project
  scope, and schematic hierarchy that reveals them;
  \item host-environment identifiers: user names, host names, absolute paths,
  license servers, and API keys; and
  \item raw simulator state and raw waveform payloads that could embed any of
  the above.
\end{itemize}
The allowed feedback is intentionally narrower than what an interactive
designer sees: scalar metrics, bounded waveform summaries, anonymized topology
intent, internal node voltages under safe aliases, and selected device
operating-point scalars drawn from a 20-key whitelist
(Section~\ref{sec:readback}).

\subsection{Attack ladder and enforcement status}

Rather than claiming blanket safety, the boundary is audited against an
explicit ladder of attack classes, summarized in Table~\ref{tab:ladder}. Tiers
1 and 2 cover direct requests for protected content and injection through
tool-call arguments (for example, SKILL expressions smuggled into identifier
fields); both are blocked by construction through the whitelist and grammar
checks of Section~\ref{sec:bridge} and are exercised by red-team unit tests.
Tier 3 covers reasoning-trace channels: content that enters a model's
reasoning trace and could be re-emitted later. Re-entry of prior reasoning
text into subsequent prompts is scrubbed at every LLM-client sink; what the
provider retains server-side is out of scope for a client-side framework and
is mitigated only by provider policy or on-premise inference. Tier 4 covers
side channels such as timing and error-class differences, which are at best
partially mitigated. We state these limits explicitly because an NDA-safe
claim is credible only when its non-goals are named.

\begin{table}[t]
  \centering
  \caption{Attack ladder for the cloud-LLM boundary and current enforcement
  status. ``Enforced'' means blocked by construction and covered by red-team
  unit tests; ``out of scope'' and ``partial'' tiers are stated limits of a
  client-side boundary and are not claimed to be red-team covered.}
  \label{tab:ladder}
  \small
  \begin{tabular}{c l l}
    \toprule
    Tier & Attack class & Status \\
    \midrule
    1 & Direct prompt-level requests for PDK/model content & enforced \\
    2 & Injection via tool arguments (SKILL/identifier breakout) & enforced \\
    3a & Reasoning-trace replay into later prompts & enforced \\
    3b & Provider-side retention of first-emission traces & out of scope \\
    4 & Timing and error-class side channels & partial \\
    \bottomrule
  \end{tabular}
\end{table}

\subsection{Hard invariants}

Five invariants are checked on every bridge call, not at construction time:
\begin{enumerate}
  \item every string returned toward the LLM passes the PDK-token, path, and
  identifier scrubber;
  \item design-variable names proposed by the LLM must match the per-project
  whitelist declared in the task specification;
  \item only whitelisted SKILL entry points may be invoked---28 at submission
  time: 25 scoped \texttt{safe*} wrappers plus three backward-compatibility
  aliases---and nested calls inside arguments are rejected except for a single
  data-constructor allow-list (\texttt{list});
  \item file uploads toward the EDA host are confined to a dedicated skill
  directory; and
  \item operating-point readback returns only keys from a 20-field whitelist
  (bias voltages, currents, small-signal parameters, threshold/overdrive,
  region, and terminal capacitances), never model-card parameters.
\end{enumerate}
These invariants are not all of one kind, and the distinction matters for what
we claim. Invariants~2, 3, and 5 are structural allowlists: they bound what the
boundary \emph{can} emit---which design variables, which SKILL entry points,
which operating-point keys---by construction, so content outside the list is
unreachable rather than merely filtered. Invariant~1, the return-path scrubber,
is different in kind: it is a conservative, deployment-specific pattern filter
that strips foundry tokens and absolute paths from free-text fields
(diagnostics, exception messages, reasoning echoes) on a best-effort basis. An
identifier that matches no seeded pattern can pass, so the scrubber is
defense-in-depth on the residual free-text surface, and the structural
allowlists---not the scrubber---carry the primary guarantee. We therefore make
the pattern list part of the deployment: a foundry extends the seeded families
with its own naming scheme before use.

One structural channel also warrants an explicit caveat. The operating-point
whitelist returns \emph{instance} operating points---per-device bias,
transconductance, and terminal capacitance---never model-card parameter names
or values. A single reading is an instance quantity, not a model coefficient;
but a provider that aggregates these scalars across the benchmark's bias points
and PVT corners could in principle regress device behaviour back toward
model-card quantities (a threshold trend, an overdrive-to-current slope). The
small per-run point count bounds but does not eliminate this inference channel;
we treat it as out of scope for a client-side boundary
(Section~\ref{sec:discussion}), and stronger parameter confidentiality is a
reason to prefer on-premise inference.

Together these invariants define what ``NDA-safe'' means in this paper: a
practical, audited scrubbing boundary under a stated adversary model, not a
formal proof of non-interference.

\section{The \framework{} Framework}
\label{sec:framework}

\subsection{Closed-loop flow}

\begin{figure}[t]
  \centering
  \includegraphics[width=\linewidth]{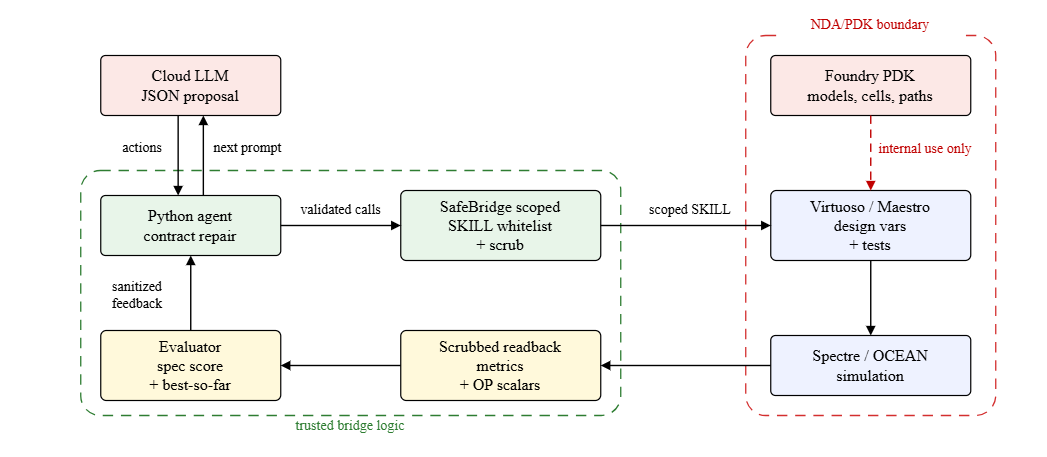}
  \caption{\framework{} data flow. The cloud LLM receives JSON actions
and scrubbed observations only. PDK content is referenced inside Cadence and
Spectre but is not serialized into model-visible feedback.}
  \Description{Data-flow diagram showing the trust boundary. The cloud LLM
  exchanges only JSON actions and scrubbed observations (topology intent,
  numeric metrics, operating-point scalars, writeback status) with the agent;
  schematic, PDK model cards, and absolute simulation paths are referenced
  inside Cadence and Spectre but never crossed back to the model.}
  \label{fig:dataflow}
\end{figure}

\framework{} is organized around a one-time setup pass followed by an iterative
optimization loop. In the setup pass, the agent first reads the target Virtuoso
schematic through \bridge{}, which returns an anonymized topology summary
(Section~\ref{sec:readback})---device roles and connectivity with foundry cell
names replaced by neutral aliases---rather than the design database itself. This
scrubbed topology is what lets the LLM understand the circuit under test and
ground the target specification, while the confidential schematic never leaves
the EDA host. Only then does the agent engage Maestro and enter the loop. Each
turn, the LLM proposes a JSON action containing design-variable updates; the
Python agent validates the action, writes allowed variables into Maestro,
triggers simulation, reads back numeric results, computes pass/fail status
against the specification, and returns a scrubbed feedback packet. The loop
stops when all hard-pass specifications are met or when one of the stop
conditions of Section~\ref{sec:contract} fires. If the iteration budget is
exhausted, the agent writes back the best numeric point observed rather than
blindly leaving the final point in Maestro. Figure~\ref{fig:dataflow} shows the
data flow and the trust boundary.

\subsection{\bridge{} enforcement layers}
\label{sec:bridge}

The bridge enforces the boundary of Section~\ref{sec:threat} in four code
locations; together with the scoped-readback filters of
Section~\ref{sec:readback} they implement the five invariants of
Section~\ref{sec:threat}. First, project specifications define the design
variables that the LLM may edit; variable names are validated against this whitelist and against a
conservative identifier grammar (letters, digits, underscore, bounded length),
and values must be single signed numeric atoms with engineering suffixes
(\texttt{f/p/n/u/m/k}); strings carrying physical units or expressions are
rejected. Second, Python validates analysis names, output expressions, and
SKILL entry points before a command reaches Virtuoso: only 28 scoped SKILL
wrappers may be named (25 \texttt{safe*} functions plus three
backward-compatibility aliases), and the full argument expression is scanned
for nested calls, which are rejected unless they belong to a small
data-constructor allow-list. Third, the SKILL wrappers on the Cadence side
expose only scoped read and write APIs (schematic readback, operating-point
readback, parameter set, OCEAN run/measure, Maestro setup and save), rather
than arbitrary database access. Fourth, every return path is scrubbed for
absolute paths, foundry-like tokens, model identifiers, and unsafe strings
before the text is reinserted into the LLM conversation; the same scrubber is
applied at every LLM-client sink, including replayed reasoning content.

\subsection{JSON action contract and stop conditions}
\label{sec:contract}

Free-form model output is not accepted into the loop. Each turn, the model
must return a single JSON object of the following shape:

\begin{lstlisting}[style=jsonstyle]
{
  "iteration": 3,
  "measurements": { "<metric_name>": <number> },
  "pass_fail":    { "<metric_name>":
       "PASS | FAIL (...) | UNMEASURABLE (...)" },
  "reasoning": "<one paragraph diagnosing the
       pass/fail pattern>",
  "design_vars": { "<var_name>":
       "<value_with_engineering_suffix>" }
}
\end{lstlisting}

The \texttt{design\_vars} keys must match the specification whitelist, and the
values must use the engineering-suffix grammar (for example \texttt{500u} or
\texttt{1.5f}); unit-bearing strings such as \texttt{mA} or \texttt{pF} are
rejected. A malformed action triggers one structured repair prompt; a second
violation terminates the run with a logged contract violation rather than
silently guessing the model's intent. The agent recomputes
\texttt{measurements} and \texttt{pass\_fail} authoritatively on the trusted
side, so a model cannot talk its way into a pass.

The loop has six machine-checked stop conditions: (i) \textsc{success}, all
hard-pass metrics pass; (ii) \textsc{max\_iter}, the iteration budget is
exhausted; (iii) \textsc{safeguard}, an oscillator-health guard fires (for the
LC-VCO task, an amplitude-hold ratio below 0.3 for three consecutive
iterations); (iv) \textsc{stuck}, identical design vectors with failing
metrics for consecutive iterations; (v) \textsc{contract\_violation}, as
above; and (vi) \textsc{no\_changes}, a first iteration that produces neither
a baseline simulation nor a proposal. Logging the terminal cause of every run
is what later makes the failure taxonomy of Section~\ref{sec:failures}
quantitative rather than anecdotal.

\subsection{Maestro setup and scoped writeback}

\framework{} treats Maestro setup as part of the reproducible experiment, not
as manual GUI state. A setup recipe can enable analyses, create output
metrics, define specs, set design variables, and verify that the requested
state actually landed under the target test row. This was necessary in
practice because AC and DC outputs need different expression families, and
because model comparison is only meaningful when every run starts from the
same documented state. The writeback path records whether test-scoped
variables were actually written; this matters for analog tasks because a
visually open Maestro session can still fail to accept writes if the wrong
setup database handle is used. At the end of a run, the best-so-far design
point (by the evaluator's score) is written back and saved, so a failed
search still leaves the database in the best observed state, and the
writeback status is part of the per-run record.

\subsection{Sanitized readback}
\label{sec:readback}

Three readback channels give the model evidence without exposing the design
database. \emph{Topology intent} converts the schematic (including a
hierarchical, same-library-only deep read) into an anonymized device-role
summary: input pair, tail source, mirrors, loads, common-mode feedback, and
output stage, with foundry cell names replaced by neutral aliases.
\emph{Metric readback} extracts AC/DC/transient and swept metrics from
Maestro/Spectre results, including multi-point sweeps for tuning curves and
per-corner scalar reads for PVT (process--voltage--temperature) sign-off. For a
PVT task the agent re-simulates the design at each named corner and returns one
scalar set per corner; the corner's supply voltage is injected through the same
trusted corner-control path used for process and temperature, never through the
LLM action space, so the model cannot edit a supply rail even while the loop
sweeps voltage. The evaluator then gates each metric on the worst corner
(worst-corner-must-pass).
\emph{Operating-point readback} returns a per-device DC table restricted to a
20-key whitelist---bias voltages (\metric{vgs}, \metric{vds}, \metric{vbs}),
currents (\metric{id}, \metric{ids}), small-signal parameters (\metric{gm},
\metric{gds}, \metric{gmb}), threshold and overdrive (\metric{vth},
\metric{vdsat}, \metric{vov}), an operating-region label, terminal
capacitances (\metric{cgs}, \metric{cgd}, \metric{cdb}, \metric{cdg},
\metric{cgg}), and per-element \metric{i}/\metric{v}/\metric{pwr} for passive
elements---so that the model can reason about bias and compensation without
ever seeing a model card.

Beyond these three channels, the framework exposes an optional,
circuit-independent hook for a deterministic, task-specific feedback module.
When a task registers one, the framework appends a bounded, ranked list of
directional design-variable suggestions that a pure-Python routine computes
from data the agent already holds. Any such module intersects its proposals
with the live design-variable whitelist and asserts the absence of foundry
tokens in its output, so it stays within the same NDA boundary as the readback
channels; it is a feedback channel, not an optimizer that bypasses the LLM. A
task that registers no module simply runs with the three channels above, and
the single instance used in this paper is described with its task in
Section~\ref{sec:protocol}.

\section{Benchmark Protocol}
\label{sec:protocol}

\subsection{Tasks}

The benchmark uses two closed-loop tasks on a proprietary 16\,nm FinFET PDK,
chosen to stress different parts of analog circuit optimization. Because the process is
FinFET, transistor sizing is set by the number of fins rather than a
width/length ratio, so the device-sizing variables in both tasks are fin
counts ($N_{\bullet}$) rather than $W/L$ values.

\begin{figure}[t]
  \centering
  \includegraphics[width=0.6\linewidth]{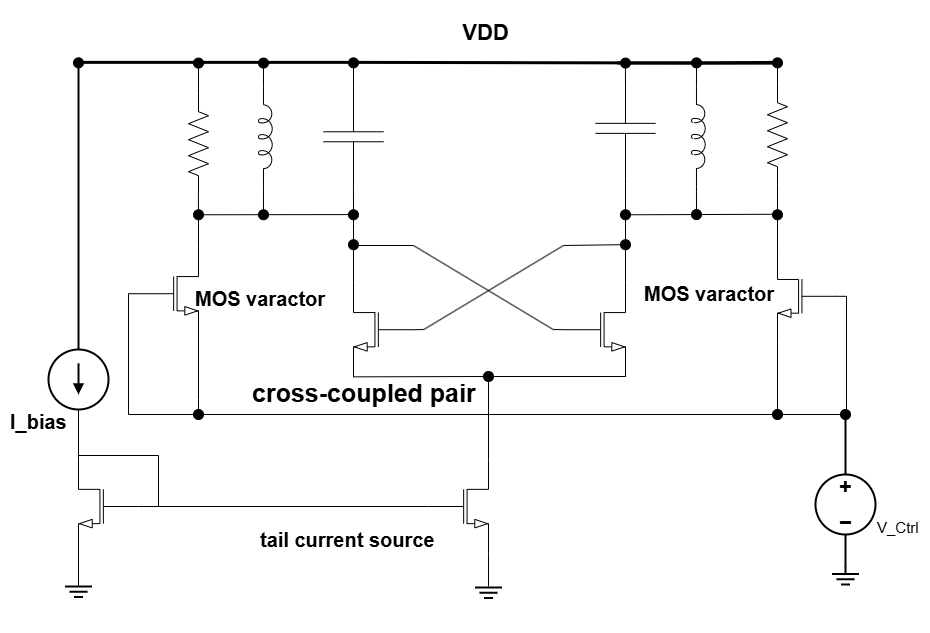}
  \caption{LC-VCO topology: a cross-coupled NMOS pair ($N_\mathrm{xc}$) resonating
with the LC tank ($L_\mathrm{tank}$, $C_\mathrm{tank}$) and a MOS varactor
($N_\mathrm{var}$) tuned by $V_{ctrl}$, biased by a mirror tail current
source ($I_\mathrm{bias}$, $N_\mathrm{tail}$). Labels match
Table~\ref{tab:lcvcovars}.}
  \Description{Schematic of an LC-VCO: a cross-coupled NMOS pair drives a parallel
  LC tank between the differential outputs and the supply, a MOS varactor tuned by
  the control voltage sets the frequency, and a mirror tail current source biases
  the core.}
  \label{fig:lcvco_sch}
\end{figure}

\textbf{LC-VCO tuning curve (PVT).} The LC-VCO task uses a 20\,GHz nominal
operating point with the control voltage fixed at $0.4$\,V ($V_{DD}/2$ at the
nominal supply) for the single-point metrics and swept over $[0.1, 0.7]$\,V
(7 points) for the tuning curve. Like the op-amp task, it is run as a
\emph{PVT} (process--voltage--temperature) sign-off: every iteration simulates
the oscillator at three corners and gates the single-point metrics on the worst
corner (worst-corner-must-pass). The three corners match the op-amp set---the
typical center (\texttt{tt}, $25\,^\circ$C, $0.80$\,V), a slow-hot corner
(\texttt{ss}, $125\,^\circ$C, $0.72$\,V), and a fast-cold corner (\texttt{ff},
$-40\,^\circ$C, $0.88$\,V)---so that the two PVT tasks share one sign-off
envelope. The pass condition combines worst-corner oscillation frequency,
differential swing, common-mode level, duty cycle, startup behavior, amplitude
hold, and core current, plus four tuning-curve metrics evaluated at the nominal
corner after the single-point worst-corner gate passes
(Table~\ref{tab:lcvcospec}). The LLM may edit eight design variables
(Table~\ref{tab:lcvcovars}; schematic in Figure~\ref{fig:lcvco_sch}); the
supply $V_{DD}$ is corner-driven through the
same trusted corner-injection path as the op-amp task and is not an LLM knob,
and the control voltage is not an optimization knob. This task stresses
oscillator transient startup across corners, swept metrics, and the need to
keep design variables physically plausible. For this
swept-metric task we enable the optional deterministic feedback channel of
Section~\ref{sec:readback}: a \emph{curve-searcher} module that, after each
tuning-curve sweep, ranks a bounded list of directional design-variable
suggestions targeting the worst-violating tuning metric. It is the only
task-specific feedback module in the benchmark; the op-amp task uses none.

\begin{table}[t]
  \centering
  \caption{LC-VCO \emph{PVT} machine-readable pass metrics. The seven
  single-point metrics are read back \emph{per corner} and gated on the worst
  of the three sign-off corners shared with the op-amp task; a run passes only
  if every band holds at every corner. The four tuning metrics are evaluated
  over the 7-point $V_{ctrl}$ sweep at the nominal corner once the single-point
  worst-corner gate passes. The $V_{cm}$ band is widened from the single-corner
  $[0.70,0.81]$ to cover the $V_{DD}$ rail at every corner, and the $K_{VCO}$
  ceiling and linearity bound are relaxed for the ideal-$LC$ concept as noted
  in the released spec.}
  \label{tab:lcvcospec}
  \small
  \begin{tabular}{l l l}
    \toprule
    Metric & Pass band & Unit \\
    \midrule
    \metric{f\_osc\_GHz} & $[19.5,\,20.5]$ & GHz \\
    \metric{V\_diff\_pp\_V} & $\geq V_{DD}/2$ & V \\
    \metric{V\_cm\_V} & $[0.65,\,0.92]$ & V (tracks $V_{DD}$) \\
    \metric{duty\_cycle\_pct} & $[48,\,52]$ & \% \\
    \metric{amp\_hold\_ratio} & $\geq 0.95$ & late/early RMS \\
    \metric{t\_startup\_ns} & $\leq 10$ & ns \\
    \metric{I\_core\_uA} & $\leq 800$ & $\mu$A \\
    \midrule
    \metric{tuning\_range\_GHz} & $\geq 0.8$ & GHz \\
    \metric{Kvco\_MHz\_per\_V} & $[150,\,3000]$ per segment & MHz/V \\
    \metric{Kvco\_linearity} & $\leq 4.0$ & max/min ratio \\
    \metric{monotonic} & true & -- \\
    \bottomrule
  \end{tabular}
\end{table}

\begin{table}[t]
  \centering
  \caption{LC-VCO design variables exposed to the LLM, with their declared
  search ranges. Symbol names are used here; the exact \texttt{design\_vars}
  keys the LLM edits are listed in the released spec.}
  \label{tab:lcvcovars}
  \small
  \begin{tabular}{l l l}
    \toprule
    Variable & Role & Range \\
    \midrule
    $I_\mathrm{bias}$ & mirror reference current & 100\,$\mu$A--2\,mA \\
    $N_\mathrm{xc}$   & cross-coupled NMOS pair (fins) & 4--32 \\
    $N_\mathrm{var}$  & MOS varactor (fins) & 4--40 \\
    $N_\mathrm{mir}$  & diode-mirror device (fins) & 4--32 \\
    $N_\mathrm{tail}$ & tail device (fins) & 8--32 \\
    $R_\mathrm{bias}$ & bias isolation resistor & 1\,k--100\,k$\Omega$ \\
    $C_\mathrm{tank}$ & tank fixed capacitor & 10\,f--200\,fF \\
    $L_\mathrm{tank}$ & tank differential inductor & 100\,p--2\,nH \\
    \bottomrule
  \end{tabular}
\end{table}

\begin{table}[t]
  \centering
  \caption{Op-amp PVT machine-readable pass metrics, evaluated on the
  differential AC testbench at $V_{icm}=V_{DD}/2$. Each metric is read back
  \emph{per corner} and gated on the worst of the three sign-off corners; a run
  passes only if every hard-pass metric holds at every corner. The phase-margin
  row was added for the PVT task to gate stability; the common-mode and CMRR
  rows are diagnostic outputs configured in the Maestro setup.}
  \label{tab:opampspec}
  \small
  \begin{tabular}{l l l}
    \toprule
    Metric & Worst-corner band & Role \\
    \midrule
    \metric{A0\_diff\_db} & $\geq 45$           & hard pass (diff.\ gain, dB) \\
    \metric{UGB\_Hz}      & $\geq 150$\,MHz      & hard pass (bandwidth) \\
    \metric{PM\_deg}      & $\geq 45$            & hard pass (phase margin, deg) \\
    \metric{Vout\_cm}     & $[0.35,\,0.45]$\,V   & DC output common mode \\
    \metric{gain\_cm\_db} & minimize             & feeds CMRR \\
    \metric{CMRR\_db}     & $\geq 40$            & ${{A0\_diff\_db}-{gain\_cm\_db}}$ (dB) \\
    \bottomrule
  \end{tabular}
\end{table}

\begin{table}[t]
  \centering
  \caption{Op-amp design variables the LLM may edit (19 sizing knobs in four
  groups). Symbol names are used here; the exact \texttt{design\_vars} keys are
  listed in the released spec. The op-amp spec declares the whitelist and the
  all-ones reset but \emph{not} per-variable ranges; the model explores within
  the \bridge{} \texttt{allowed\_params} whitelist and physical plausibility.
  The four stimulus variables (input common mode, AC magnitude, two AC phases)
  are held fixed and excluded.}
  \label{tab:opampvars}
  \small
  \begin{tabular}{@{}l l@{}}
    \toprule
    Group (count) & Variables ($N$: fin counts, $I$: currents, $R$/$C$) \\
    \midrule
    Differential-stage sizing (7) &
      $N_\mathrm{n1},N_\mathrm{n2},N_\mathrm{p1},N_\mathrm{p2},N_\mathrm{nb},N_\mathrm{pb},N_\mathrm{tail}$ \\
    Bias currents (2) & $I_\mathrm{n},I_\mathrm{p}$ \\
    CMFB (7) &
      $N_\mathrm{cn1},N_\mathrm{cn2},N_\mathrm{cp1},N_\mathrm{cp2},N_\mathrm{cb},N_\mathrm{cc},I_\mathrm{c}$ \\
    Miller \& CM sense (3) & $R_\mathrm{m},C_\mathrm{m},R_\mathrm{cm}$ \\
    \bottomrule
  \end{tabular}
\end{table}

\begin{figure}[t]
  \centering
  \includegraphics[width=0.62\linewidth]{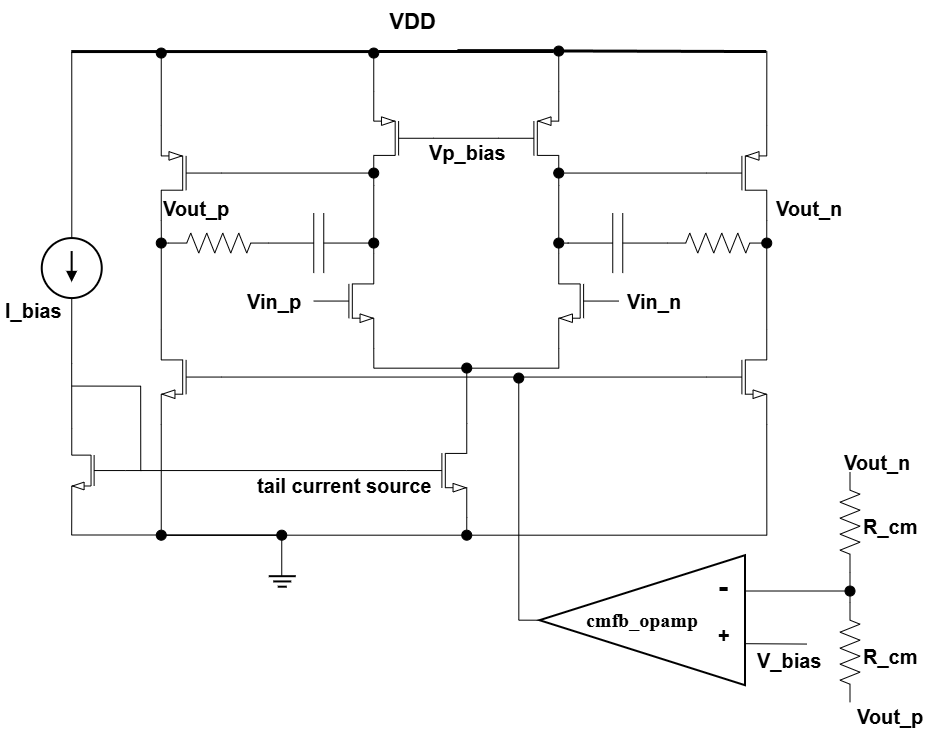}
    \caption{Two-stage Miller-compensated op-amp (single-ended core shown): an NMOS
input pair ($N_\mathrm{n1}$) with a PMOS mirror load, a common-source second
stage ($N_\mathrm{p1}$), and Miller compensation ($R_\mathrm{m}$,
$C_\mathrm{m}$). The benchmark circuit is the fully differential version of this
core, with a resistor-sensed common-mode feedback ($R_\mathrm{cm}$) loop setting
the output common mode. Labels match the groups of Table~\ref{tab:opampvars}.}
  \Description{Schematic of a two-stage Miller-compensated op-amp: an NMOS input
  differential pair with a PMOS mirror load forms the first stage, a common-source
  PMOS second stage drives the output, and a series resistor-capacitor Miller
  network compensates between the two stages.}
    \label{fig:opamp_sch}
\end{figure}

\textbf{Two-stage op-amp (PVT).} The op-amp task uses a differential AC
testbench with common-mode input stimulus, output load capacitance, and a fully
differential two-stage op-amp with resistor-sensed common-mode feedback and
Miller RC compensation (Figure~\ref{fig:opamp_sch}). It is run as a \emph{PVT}
(process--voltage--temperature) sign-off task: every iteration simulates the
design at three corners and gates each AC metric on the worst corner
(worst-corner-must-pass). The three corners are the typical center (\texttt{tt},
$25\,^\circ$C, $0.80$\,V), a slow-hot corner (\texttt{ss}, $125\,^\circ$C,
$0.72$\,V), and a fast-cold corner (\texttt{ff}, $-40\,^\circ$C, $0.88$\,V),
spanning all three process labels, both temperature extremes, and the
$\pm 10\%$ supply range. The two most extreme corners of a full five-corner set
(slow/cold/low-supply and fast/hot/high-supply) are excluded from sign-off
based on exploratory five-corner runs: no sizing point within the declared
variable ranges held gain at the slow/cold/$0.72$\,V corner, where the
push-pull second stage falls out of its operating region, so retaining those
corners would make the task unsatisfiable by construction rather than
discriminating. The three retained corners form the feasible PVT envelope. The
hard-pass targets are a worst-corner differential gain $A_0 \ge 45$\,dB, a
worst-corner unity-gain bandwidth $\ge 150$\,MHz, and---added for the PVT
task---a worst-corner phase margin $\ge 45^\circ$ that gates stability so a
high-gain but under-compensated point cannot pass. The worst-corner gain bound
is $45$\,dB rather than the $50$\,dB single-corner target because the worst of
three corners naturally sits a few dB below the typical corner. The LLM may edit
nineteen sizing variables (stage and CMFB device sizes, bias currents, Miller
$R$ and $C$, and the common-mode sense resistor); the supply $V_{DD}$ is
\emph{not} LLM-editable---it is driven per corner through a trusted
corner-injection path that keeps supply rails out of the LLM action space while
still sweeping voltage. The full per-metric bands are listed in
Table~\ref{tab:opampspec}. The four stimulus variables (input common mode, AC
magnitude, and the two AC phases) are held fixed by the harness so that a model
cannot pass by editing the testbench instead of the circuit.

\subsection{Fair-start protocol}

Each model is evaluated from the same reset state. For the op-amp, all
non-stimulus sizing variables are initialized to one---a deliberately
unbiased, nearly pathological start---while stimulus variables use the
differential AC preset (common-mode input $V_{DD}/2$, antiphase AC drive with
0/180 degree phases). For the LC-VCO, all sizing variables likewise start
from the documented all-ones reset, the supply is driven per corner through the
trusted corner path, and the control voltage is fixed by the nominal and sweep
definitions. The reset is applied programmatically through
the Maestro writeback path and verified before the first iteration; each
per-run record stores the reset report. The iteration budget is 15 unless
otherwise specified. A run is counted as \pass{} only if all hard-pass
metrics are satisfied before the budget ends. Every run logs its terminal
cause (success, metric miss at budget, contract violation, LLM/client error,
or harness failure), wall-clock time, iteration count, final and best metrics,
final design vector, and writeback status, as CSV/JSONL artifacts.

The main tables report eleven matched model checkpoints spanning seven vendor
families. The same model set is used for the op-amp and LC-VCO tasks so that
cross-task comparisons are not confounded by different endpoint coverage. Each
matched cell is a single audited run (seed~1); we discuss the implications of
single-seed snapshots in Section~\ref{sec:discussion}.

\section{Results}
\label{sec:results}

\subsection{Overall outcome}

Both tasks are run under the same \emph{PVT}
(process--voltage--temperature) sign-off discipline, with worst-corner gating
across the same three corners, so their pass rates reflect the circuits'
intrinsic closure difficulty rather than any asymmetry in the test regime. On the LC-VCO task, 7 of 11 models pass, with
passing runs requiring four to fifteen iterations (Table~\ref{tab:lcvco}). The
op-amp task is the harder of the two: only 4 of 11 models pass
(Table~\ref{tab:opamp}). The failure signatures explain the gap. LC-VCO
failures are one-dimensional---the worst-corner oscillation frequency falls
outside the $[19.5,20.5]$\,GHz band (an under-frequency miss for
claude-sonnet-4-6, over-frequency for gpt-5.4-mini, or no sustained
oscillation at all for claude-haiku-4-5)---so a failing model misses on
frequency placement alone. The fourth LC-VCO failure is not a circuit miss at
all: minimax-m2.7 terminated through an unrecoverable client error at
iteration~7 (Section~\ref{sec:failures}), so its row measures the endpoint
stack rather than the design. The op-amp failures are multi-dimensional:
several models never reach the worst-corner gain target ($A_0\!<\!45$\,dB at
that corner) or collapse the design entirely, while two models (claude-opus-4-8 and
gpt-5.4-mini) clear the worst-corner gain and bandwidth gates but with a
\emph{negative} worst-corner phase margin---a high-gain operating point that
would oscillate, which the phase-margin gate rejects as unstable. Holding gain,
bandwidth, \emph{and} stability simultaneously across three corners is a
substantially tighter closure condition than aligning a single oscillation
frequency, and it exposes that the loop can optimize a visible metric while
missing the coupled biasing and compensation tradeoffs required for full
specification closure.

\begin{table*}[t]
  \centering
  \caption{Eleven-model LC-VCO \emph{PVT} benchmark. Each single-point metric
  is gated on the worst of the same three sign-off corners as the op-amp task
  (\texttt{tt}/$25\,^\circ$C/$0.80$\,V, \texttt{ss}/$125\,^\circ$C/$0.72$\,V,
  \texttt{ff}/$-40\,^\circ$C/$0.88$\,V); the $f_{osc}$, $V_{pp}$, and
  $I_{core}$ columns report the worst-corner value, and the tuning-range and
  $K_{VCO}$-linearity columns are read at the nominal corner after the
  single-point gate passes (blank for runs that never reached the sweep phase).
  All runs start from the same documented reset state and use a 15-iteration
  budget. The model set matches the op-amp benchmark exactly.}
  \label{tab:lcvco}
  \scriptsize
  \setlength{\tabcolsep}{3.5pt}
  \begin{tabular}{l c r r r r r r r l}
    \toprule
    Model & Outcome & Iter. & Time (s) & $f_{osc}$ (GHz) & $V_{pp}$ (V) &
    $I_{core}$ ($\mu$A) & Range (GHz) & $K_{VCO}$ lin. & Failure reason \\
    \midrule
    claude-opus-4-8   & \pass &  5 & 2493 & 20.06 & 3.28 & 359.9 & 0.824 & 2.06 & -- \\
    claude-sonnet-4-6 & \fail & 15 & 4093 & 19.33 & 3.14 & 473.0 & --    & --   & max\_iter ($f_{osc}\!<\!19.5$) \\
    claude-haiku-4-5  & \fail & 15 & 3709 & --    & --   & --    & --    & --   & max\_iter (no osc.) \\
    gpt-5.5           & \pass &  6 & 3404 & 19.94 & 3.28 & 359.6 & 0.833 & 1.99 & -- \\
    gpt-5.4-mini      & \fail & 15 & 3912 & 20.52 & 3.09 & 415.6 & --    & --   & max\_iter ($f_{osc}\!>\!20.5$) \\
    kimi-k2.5         & \pass &  4 & 2165 & 20.00 & 3.26 & 365.3 & 1.005 & 1.95 & -- \\
    minimax-m2.7      & \fail &  7 & 2523 & 20.86 & 3.18 & 372.4 & --    & --   & llm\_error \\
    minimax-m3        & \pass &  6 & 3866 & 19.93 & 3.26 & 361.4 & 1.076 & 1.93 & -- \\
    deepseek-v4-pro   & \pass &  8 & 3483 & 20.15 & 3.28 & 361.7 & 0.845 & 2.11 & -- \\
    deepseek-v4-flash & \pass &  6 & 2344 & 20.00 & 3.19 & 365.6 & 1.331 & 3.44 & -- \\
    gemini-2.5-pro    & \pass & 15 & 5912 & 20.07 & 3.26 & 365.8 & 1.165 & 2.85 & -- \\
    \bottomrule
  \end{tabular}

  \smallskip
  \raggedright\footnotesize The three LC-VCO metric-miss failures all miss the
  same hard metric---worst-corner $f_{osc}$ outside $[19.5,20.5]$\,GHz---and
  terminate at the iteration budget: claude-sonnet-4-6 stalls under-frequency
  (19.33\,GHz), gpt-5.4-mini over-frequency (20.52\,GHz), and claude-haiku-4-5
  never sustains oscillation at the worst corner (no measurable $f_{osc}$).
  minimax-m2.7 terminated early through an \metric{llm\_error}---an
  unrecoverable client error at iteration~7, not a metric miss. Because the
  single-point worst-corner gate precedes the tuning sweep, none of the four
  failing runs reached the sweep phase, so their tuning-range and $K_{VCO}$
  columns are blank.
\end{table*}

\begin{table*}[t]
  \centering
  \caption{Op-amp \emph{PVT} benchmark. Each metric is gated on the worst of
  three sign-off corners---typical (\texttt{tt}, $25\,^\circ$C, $0.80$\,V),
  slow-hot (\texttt{ss}, $125\,^\circ$C, $0.72$\,V), and fast-cold (\texttt{ff},
  $-40\,^\circ$C, $0.88$\,V); the $A_0$, UGB, and phase-margin (PM) columns
  report the worst-corner value. A run is \pass{} only if all three hard gates
  ($A_0\!\ge\!45$\,dB, $\mathrm{UGB}\!\ge\!150$\,MHz, $\mathrm{PM}\!\ge\!45^\circ$)
  hold at every corner. All runs use the 15-iteration budget and the same model
  set as the LC-VCO benchmark.}
  \label{tab:opamp}
  \footnotesize
  \begin{tabular}{l c r r r r r l}
    \toprule
    Model & Outcome & Iter. & Time (s) & $A_0$ (dB) & UGB (MHz) & PM (deg) & Failure reason \\
    \midrule
    claude-opus-4-8   & \fail & 15 & 1663 & 51.76    & 695.0  & $-6.78$  & PM$<45$ (unstable)$^{\ddagger}$ \\
    claude-sonnet-4-6 & \pass &  9 & 1144 & 50.35    & 257.0  & 150.33   & -- \\
    claude-haiku-4-5  & \fail & 15 & 1785 & 39.88    & 1490.9 & 28.47    & max\_iter ($A_0$, PM) \\
    gpt-5.5           & \pass &  3 &  397 & 52.53    & 212.9  & 101.51   & -- \\
    gpt-5.4-mini      & \fail & 15 & 1467 & 41.05    & 396.5  & $-36.12$ & PM$<45$ (unstable)$^{\ddagger}$ \\
    kimi-k2.5         & \fail & 15 & 2645 & $-13.32$ & 114.6  & --       & max\_iter \\
    minimax-m2.7      & \pass &  9 & 1305 & 49.50    & 313.0  & 108.39   & -- \\
    minimax-m3        & \fail &  2 & 1810 & $-9.45$  & 2496.3 & $-85.96$ & contract\_violation \\
    deepseek-v4-pro   & \pass &  9 & 1942 & 50.46    & 848.2  & 89.55    & -- \\
    deepseek-v4-flash & \fail & 15 & 2194 & 37.10    & 710.9  & 137.11   & max\_iter ($A_0$) \\
    gemini-2.5-pro    & \fail & 15 & 1807 & $-74.82$ & --     & 70.13    & max\_iter \\
    \bottomrule
  \end{tabular}

  \smallskip
  \raggedright\footnotesize $^{\ddagger}$The design clears the worst-corner
  $A_0$ and UGB gates but its worst-corner phase margin is negative---a
  high-gain operating point that would oscillate. The phase-margin gate added
  for the PVT task rejects it as an unstable (non-converged) design rather than
  a pass; without it, $A_0$/UGB alone would have falsely counted these two runs
  as passes.
\end{table*}

\subsection{Tuning-curve quality of passing runs}

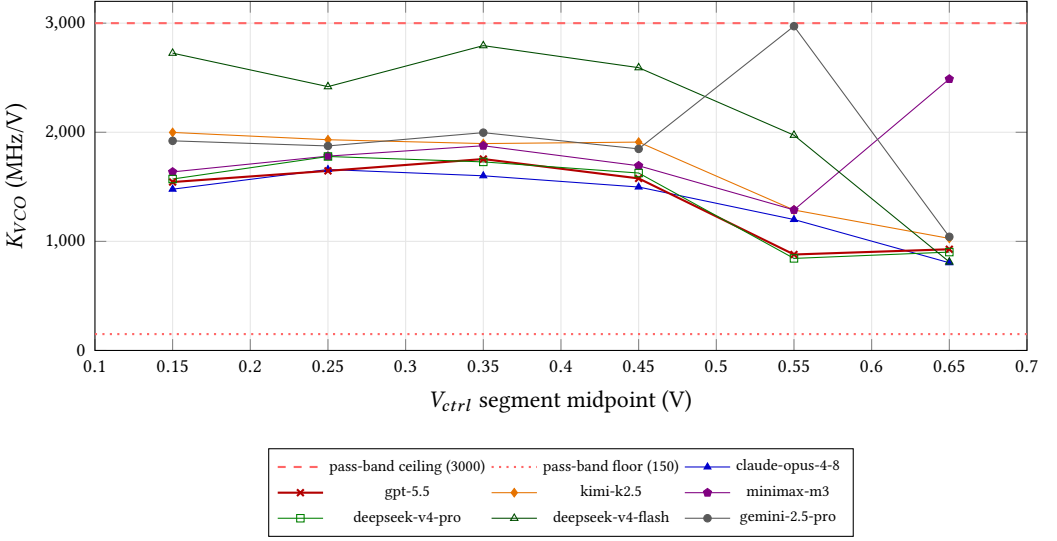
\begin{figure}[t]
\centering
\begin{tikzpicture}
\begin{axis}[
  width=\linewidth,
  height=6.2cm,
  xlabel={$V_{ctrl}$ segment midpoint (V)},
  ylabel={$K_{VCO}$ (MHz/V)},
  xmin=0.1, xmax=0.7,
  ymin=0, ymax=3200,
  legend style={font=\tiny, at={(0.5,-0.28)}, anchor=north, legend columns=3},
  grid=major,
  grid style={gray!20},
  tick label style={font=\scriptsize},
  label style={font=\small},
]
\addplot[red!60, dashed, thick, domain=0.1:0.7] {3000};
\addlegendentry{pass-band ceiling (3000)}
\addplot[red!60, dotted, thick, domain=0.1:0.7] {150};
\addlegendentry{pass-band floor (150)}
\addplot[blue!80!black, mark=triangle*, mark size=1.5] coordinates
  {(0.15,1478.2) (0.25,1658.6) (0.35,1601.0) (0.45,1498.1) (0.55,1200.2) (0.65,803.8)};
\addlegendentry{claude-opus-4-8}
\addplot[red!70!black, mark=x, mark size=1.8, thick] coordinates
  {(0.15,1543.7) (0.25,1646.0) (0.35,1754.5) (0.45,1577.0) (0.55,879.6) (0.65,927.5)};
\addlegendentry{gpt-5.5}
\addplot[orange!90!black, mark=diamond*, mark size=1.6] coordinates
  {(0.15,1998.5) (0.25,1931.5) (0.35,1896.3) (0.45,1909.8) (0.55,1287.4) (0.65,1026.9)};
\addlegendentry{kimi-k2.5}
\addplot[violet, mark=pentagon*, mark size=1.6] coordinates
  {(0.15,1636.6) (0.25,1781.3) (0.35,1876.4) (0.45,1693.0) (0.55,1288.3) (0.65,2488.2)};
\addlegendentry{minimax-m3}
\addplot[green!50!black, mark=square, mark size=1.4] coordinates
  {(0.15,1571.2) (0.25,1778.0) (0.35,1728.4) (0.45,1626.4) (0.55,843.7) (0.65,901.7)};
\addlegendentry{deepseek-v4-pro}
\addplot[green!30!black, mark=triangle, mark size=1.6] coordinates
  {(0.15,2725.2) (0.25,2418.2) (0.35,2794.1) (0.45,2592.2) (0.55,1973.1) (0.65,811.5)};
\addlegendentry{deepseek-v4-flash}
\addplot[gray!70!black, mark=*, mark size=1.4] coordinates
  {(0.15,1921.3) (0.25,1874.7) (0.35,1996.4) (0.45,1847.6) (0.55,2972.2) (0.65,1041.8)};
\addlegendentry{gemini-2.5-pro}
\end{axis}
\end{tikzpicture}
\caption{Per-segment $K_{VCO}$ of each passing model's final LC-VCO design
point over the $V_{ctrl}$ sweep at the nominal corner, from the benchmark
artifacts. All seven passing models stay inside the $[150, 3000]$\,MHz/V
per-segment band (dashed/dotted red). The four failing runs are absent because
they never cleared the single-point worst-corner $f_{osc}$ gate that precedes
the tuning sweep.}
\Description{Line plot of per-segment VCO gain (K-VCO, MHz/V) versus the control
voltage segment midpoint for every passing model's final design. All seven
passing models stay within the 150 to 3000 MHz/V pass band.}
\label{fig:kvco}
\end{figure}

Pass/fail alone hides how differently the models size the same oscillator.
Figure~\ref{fig:kvco} plots the per-segment $K_{VCO}$ of each passing model's
final design point across the $V_{ctrl}$ sweep at the nominal corner, against
the $[150, 3000]$\,MHz/V per-segment pass band. The seven passing models land
tuning curves of similar shape---monotonically decreasing gain toward the top
of the varactor range---but with visibly different flatness, which the
\metric{Kvco\_linearity} column of Table~\ref{tab:lcvco} quantifies (1.93 for
minimax-m3 and 1.95 for kimi-k2.5 at the flat end versus 3.44 for
deepseek-v4-flash at the steep end). The four failing runs never reach this
plot: because the tuning sweep runs only after the single-point worst-corner
gate passes, and all four miss the worst-corner $f_{osc}$ band (or, for
minimax-m2.7, terminate on a client error first), none produced a tuning curve.
The LC-VCO difficulty under PVT is therefore concentrated in placing the
worst-corner oscillation frequency inside the band: every model that clears
that gate also lands a tuning curve inside the task's per-segment bounds. That
second observation should be read against the calibrated bands of
Table~\ref{tab:lcvcospec}---under the base variant's tighter 2000\,MHz/V
ceiling, three of the seven passing curves (minimax-m3, deepseek-v4-flash,
gemini-2.5-pro) would have exceeded the per-segment limit---so the relaxed
ceiling shifts the discriminating gate to frequency placement rather than
eliminating tuning-curve difficulty outright.

\subsection{Cross-task observations}

First, passing one topology does not imply passing the other, even under the
same PVT gate. Several strong checkpoints pass both tasks, but the mismatches
remain informative and now run in both directions. deepseek-v4-flash passes
LC-VCO while failing the op-amp PVT loop with a worst-corner gain miss at the
iteration budget; conversely, claude-sonnet-4-6 and minimax-m2.7 pass the
op-amp PVT loop while missing LC-VCO---sonnet stalling under-frequency and
minimax-m2.7 lost to a client error. gpt-5.4-mini misses both tasks with
different signatures: an unstable op-amp point (negative worst-corner phase
margin) and an over-frequency LC-VCO point at the budget. Claude-haiku-4-5 also
misses both, its LC-VCO row a worst-corner metric miss (no sustained
oscillation) rather than a tool failure. This bidirectional split supports
treating analog agent evaluation as a multi-topology benchmark rather than a
single-circuit leaderboard.

Second, the op-amp failures are informative rather than random. Several failed
runs produce very high worst-corner UGB values but insufficient or negative
low-frequency gain, or---the failure mode the PVT phase-margin gate newly
exposes---high gain and bandwidth but a negative worst-corner phase margin.
Both are recognizable analog-design failure modes: the model has found a fast
or misbiased point without the operating-region and compensation balance needed
for gain, or it has traded compensation away for gain and produced an unstable
amplifier. The framework changes these from an opaque model failure into a
debuggable trace because the LLM-visible feedback contains per-corner AC
metrics (now including phase margin), DC readback, topology intent, and
operating-point scalars.

Third, tool-protocol reliability is part of model quality in an EDA agent. On
the op-amp PVT task, minimax-m3 terminated through a
\metric{contract\_violation} after two iterations: it repeatedly returned JSON
that omitted required action keys, and the harness repair policy could not
recover a well-formed action. This is not an analog metric miss but a
tool-contract failure, and it exposes an important deployment risk: a model
that can reason about circuits but cannot reliably emit the contracted action
loop is not deployable as a closed-loop EDA agent. A deployable agent must
therefore be evaluated at the level of the complete tool loop, including the
provider endpoint, response contract, parser, and repair policy; rankings must
be read as dated snapshots of the served model plus its client stack.

Fourth, iteration efficiency varies substantially among passing models. On
LC-VCO, kimi-k2.5 passes in four iterations (2165\,s wall clock), while
gemini-2.5-pro needs the full fifteen (5912\,s). The snapshot also shows that a
pass can arrive after the same iteration count at very different wall-clock
cost: gpt-5.5, minimax-m3, and deepseek-v4-flash all pass in six iterations,
but at 3404\,s, 3866\,s, and 2344\,s respectively. On the op-amp PVT task,
gpt-5.5 passes in three iterations (397\,s), whereas claude-sonnet-4-6,
minimax-m2.7, and deepseek-v4-pro each pass at nine iterations but at
1144\,s, 1305\,s, and 1942\,s respectively---the same iteration count over
three corners at very different wall-clock cost. Since every iteration now
costs a per-corner Spectre run at three corners plus model tokens on both
tasks, iteration count is a first-order proxy for deployment cost, and the
per-run wall-clock and iteration columns in Tables~\ref{tab:lcvco} and
\ref{tab:opamp} are part of the published artifacts.

\section{Feedback-Path Case Ablations}
\label{sec:ablation}

Each sanitized feedback channel is both optimization evidence and audit
surface, so a natural question is which channels actually change outcomes. As
preliminary, design-motivating evidence---a single mid-tier checkpoint
(mimo-v2.5-pro), single seed, 5-iteration budget, from the same all-ones reset
on the single-corner base LC-VCO variant of the task (the feedback-channel
mechanism under test is identical to the PVT benchmark; only the corner set and
the $K_{VCO}$ ceiling differ)---we disabled feedback paths one at a time
(Table~\ref{tab:ablation}). We report these because each directly justifies a
framework design decision; they are case observations, not statistical claims.

\begin{table}[t]
  \centering
  \caption{Single-corner base LC-VCO feedback-path ablations for mimo-v2.5-pro
  (single seed, 5-iteration evidence budget; the $K_{VCO}$ ceiling here is the
  base variant's 2000\,MHz/V, not the PVT benchmark's 3000\,MHz/V).
  ``Nominal-only \pass{}'' means all evaluated metrics passed but the
  tuning-curve metrics were never evaluated because the swept-metric path was
  disabled.}
  \label{tab:ablation}
  \small
  \begin{tabular}{l c r l}
    \toprule
    Variant & Outcome & Iter. & Note \\
    \midrule
    full              & \pass & 3 & range 0.887\,GHz, $K_{VCO}$ lin.\ 1.99 \\
    no swept metrics  & nominal-only \pass & 2 & tuning spec silently unevaluated \\
    no curve searcher & \fail & 5 & $K_{VCO}$ segments above 2000\,MHz/V \\
    no writeback      & \pass & 4 & final state not persisted to Maestro \\
    \bottomrule
  \end{tabular}
\end{table}

The first decision-relevant result is that disabling the swept-metric path does
\emph{not} degrade gracefully: the run still reports \pass{} in fewer
iterations, because the tuning-curve portion of the specification silently
drops out of the evaluation. In a safety-bounded loop a feedback ablation can
thus masquerade as an improvement---which is exactly why the framework records
the \emph{evaluated metric set}, not just the pass verdict, as part of every run
record. The second is that removing the deterministic curve-searcher turns the
same task into a failure within the budget: the model widens the tuning range
but leaves per-segment $K_{VCO}$ above the base variant's 2000\,MHz/V
ceiling---direct evidence that structured, PDK-safe directional hints are cheap
to sanitize and measurably help hold the tuning-curve specification. (Disabling
best-so-far writeback did not change this run's verdict, consistent with its
role as state preservation rather than search assistance; its value shows up in
failed runs, where the database would otherwise be left at an arbitrary point.)

A companion full-feedback sweep over seven other checkpoints at the same
5-iteration budget broadly tracks the main table but is budget-sensitive:
deepseek-v4-flash fails here yet passes under the 15-iteration matched budget,
confirming that the iteration budget materially affects rankings even when the
feedback path is unchanged. Ablations of the remaining
channels---topology intent, operating-point scalars, and the strict JSON
protocol versus a loose natural-language variant---across all checkpoints and
multiple seeds are ongoing work and are not claimed here.

More broadly, because model quality varies sharply across the matched set (seven
of eleven pass the LC-VCO PVT task, with the failures showing distinct
signatures)
and removing individual feedback channels changes outcomes as shown above, the
loop's result is attributable to the model's reasoning rather than to the fixed
scaffolding---the simulator, the deterministic curve-searcher, or the all-ones
reset---acting alone.
\section{Failure Analysis}
\label{sec:failures}

\subsection{Observed terminal causes}

Because every run logs a machine-readable terminal cause, the failure
analysis can start from counts rather than anecdotes.
Table~\ref{tab:failcounts} aggregates the 22 matched runs reported in
Section~\ref{sec:results} (11 LC-VCO, 11 op-amp).

\begin{table}[t]
  \centering
  \caption{Terminal causes over the 22 matched benchmark runs of
  Section~\ref{sec:results}.}
  \label{tab:failcounts}
  \small
  \begin{tabular}{l r l}
    \toprule
    Terminal cause & Runs & Instances \\
    \midrule
    all hard-pass metrics met & 11 & 7 LC-VCO, 4 op-amp \\
    metric miss at budget & 9 & 3 LC-VCO, 6 op-amp \\
    LLM/client error & 1 & 1 LC-VCO \\
    contract violation & 1 & 1 op-amp \\
    \bottomrule
  \end{tabular}
\end{table}

\subsection{Failure taxonomy}
\label{sec:taxonomy}

The taxonomy below classifies how closed-loop analog agents break. Buckets
1 and 2 were observed as terminal causes or failure signatures in the matched
runs; bucket 4 was the terminal cause of one matched op-amp run (minimax-m3),
and bucket 5 the terminal cause of one matched LC-VCO run (minimax-m2.7) in
addition to a transient provider outage that the Anthropic checkpoints recovered
from before completion; buckets 3, 6, 7, and 8 are boundary- or
robustness-guarded behaviors, exercised by repair prompts and red-team unit
tests rather than occurring as terminal causes in this snapshot.

\begin{enumerate}
  \item \textbf{Metric miss at budget}: simulation completes but a hard-pass
  metric is never satisfied at the worst corner, such as op-amp worst-corner
  gain stalling below 45\,dB (claude-haiku-4-5 at 39.88\,dB, deepseek-v4-flash
  at 37.10\,dB) or LC-VCO worst-corner oscillation frequency stuck outside the
  $[19.5,20.5]$\,GHz band at the iteration budget---under-frequency for
  claude-sonnet-4-6 (19.33\,GHz), over-frequency for gpt-5.4-mini (20.52\,GHz),
  or no sustained oscillation at all for claude-haiku-4-5. A new sub-case under
  PVT is a \emph{stability} miss: claude-opus-4-8 and gpt-5.4-mini reach the
  worst-corner gain and bandwidth gates but with a negative worst-corner phase
  margin, so the phase-margin gate fails them as unstable high-gain points.
  \item \textbf{Bias or operating-region miss}: a sub-class of the above with
  a recognizable signature---high worst-corner bandwidth with low or negative
  worst-corner gain (for example, gemini-2.5-pro collapses to a negative
  worst-corner gain and kimi-k2.5 reaches only $-13.3$\,dB worst-corner gain),
  indicating a misbiased operating point rather than a near-miss.
  \item \textbf{Identical-vector stall}: the model repeatedly proposes the
  same failing design vector after receiving valid measurement feedback. The
  identical-vector stall guard (Section~\ref{sec:contract}) exists to terminate
  such runs, but it was not the terminal cause of any run in this matched PVT
  snapshot; each metric-miss run instead ran to the iteration budget.
  \item \textbf{Contract failure} (protocol-enforced): the model cannot hold
  the JSON action contract; after the structured repair attempts, the run
  terminates with a logged violation (Section~\ref{sec:contract}). On the op-amp
  PVT task this was the terminal cause for minimax-m3, which repeatedly returned
  actions missing required keys and could not be repaired into a well-formed
  action.
  \item \textbf{LLM/client loop error}: the model API, streaming client,
  parser, or repair loop returns an unrecoverable error before a metric-based
  terminal condition is reached. This was the terminal cause of one matched
  LC-VCO run: minimax-m2.7 aborted with an \metric{llm\_error} at iteration~7
  after producing a valid but out-of-band point (20.86\,GHz worst corner), its
  client returning an unrecoverable error before the budget. The same class also
  appears as a transient: an earlier provider outage during the PVT campaign
  caused the three Anthropic checkpoints to abort with a client-side error
  before producing any metric, and they only completed the matched run after the
  endpoint recovered. This bucket is a property of the served model plus client
  stack at benchmark time, not of the circuit alone.
  \item \textbf{Harness failure}: the benchmark runner terminates before a
  final result is parsed---a framework-level failure rather than a model
  failure. The framework's data-integrity safeguards (Section~\ref{sec:repro})
  keep this class out of the matched terminal causes.
  \item \textbf{PDK-content probes} (enforced): tool calls that request
  foundry content by name or via SKILL injection are blocked by the
  entry-point whitelist and scrubber (Tiers 1--2 of Table~\ref{tab:ladder}).
  \item \textbf{Reasoning-trace replay} (enforced): prior reasoning content
  re-entering the prompt passes through the same scrubber as tool results
  (Tier 3a of Table~\ref{tab:ladder}).
\end{enumerate}
The remaining degenerate-search guards---oscillator-death safeguard and
unmeasurable-metric saturation---exist in the protocol
(Section~\ref{sec:contract}) but were not terminal causes in this snapshot.

\subsection{What the failures say about feedback design}

The dominant real failure mode is not protocol chaos but \emph{partial}
optimization: models reliably improve the metrics they can see moving
(bandwidth, range) and stall on metrics that require coupled, multi-variable
reasoning (gain at a healthy bias point, per-segment gain flatness). This is
precisely where sanitized operating-point evidence and structured directional
hints earn their audit cost, as the ablation of Section~\ref{sec:ablation}
indicates. It also explains why the same model family can split across tasks:
the LC-VCO loop exposes a curve-level summary after every sweep, while the
op-amp loop exposes per-corner scalar AC metrics (now including phase margin)
plus the DC table, leaving more of the cross-corner compensation reasoning to
the model.

\section{Reproducibility and Security Audit}
\label{sec:repro}

The reproducibility package has two layers. The public layer is released as an
open-source repository%
\ifanon
\else
\footnote{\url{https://github.com/lixunqi12/virtuoso-agent}}%
\fi
{} and contains the agent code, project specifications, Maestro setup recipes,
scrubbed transcripts, benchmark summaries (CSV/JSONL, including every value in
Tables~\ref{tab:lcvco}--\ref{tab:failcounts} and
Figure~\ref{fig:kvco}), and leak-check scripts. The private layer is supplied
by the user: a licensed Virtuoso/Maestro/Spectre installation, an opened
Maestro session for the target testbench, and a foundry PDK that remains
local. This split is intentional: a reviewer can inspect the safety mechanism
and replay scrubbed benchmark evidence without receiving the PDK, and a
licensed user can re-run the full loop without any artifact from us touching
their NDA boundary.

The security audit checks three classes of leakage. First, source-level leak
checks scan the repository for private paths, API keys, remote host names,
and foundry-shaped tokens; the gate run accompanying this draft passes with
no sensitive patterns in the PC-side source workspace. Second,
bridge-level tests verify that schematic readback, operating-point readback,
OCEAN result generation, and Maestro writeback return only whitelisted
fields, and that non-whitelisted SKILL entry points and nested-call
injections are rejected. Third, transcript-level checks verify that the
exact text sent back to the LLM does not include PDK model names, absolute
paths, or replayed reasoning content that bypasses the scrubber. The
benchmark artifacts published with the paper are the same files the gate
scans, so the audit covers the evidence, not only the code.

Beyond these static checks, we measure leak resistance under \emph{active}
attack. A leak oracle scans the final text returned to the LLM---the assembled
feedback prompt, not the bridge return object---against the production
scrub-pattern gate plus a registry of planted synthetic canaries, one per
protected-asset class. Three attacker modalities exercise it
(Table~\ref{tab:redteam}): a fixed, deterministic probe suite that drives the
real readback and injection paths with canary-laden input; a live readback of
the two real benchmark schematics through the deployed bridge; and an adaptive
LLM red-teamer (deepseek-v4-pro) that proposes a fresh injection each turn over
eight-turn sessions, adapting to what the bridge returns. Across all three---%
eight fixed probes, two live cell readbacks, and ten adaptive sessions---no
protected token reached the model (attack-success rate~0). The measurement is
token/sentinel-level under the stated threat model, not a non-interference
proof; consistent with Section~\ref{sec:threat}, path redaction is keyed to a
known-root allowlist that covers the deployment's \texttt{/project}-rooted
paths.

\begin{table}[t]
  \centering
  \caption{Adversarial extraction results. A trial \emph{leaks} if the oracle
  (the production scrub-pattern gate plus a planted synthetic-canary registry)
  matches a protected token in the final text returned to the LLM. Fixed probes
  are counted per-probe; the adaptive LLM attacker (deepseek-v4-pro, eight turns
  per session) per-session.}
  \label{tab:redteam}
  \small
  \begin{tabular}{@{}llccc@{}}
    \toprule
    Attacker & Surface / tier & Trials & Leaks & ASR \\
    \midrule
    Fixed probe   & T1 direct readback        & 4  & 0 & 0\% \\
    Fixed probe   & T2 tool-arg injection      & 2  & 0 & 0\% \\
    Fixed probe   & T3a reasoning replay       & 2  & 0 & 0\% \\
    \midrule
    Live readback & real LC-VCO + op-amp       & 2  & 0 & 0\% \\
    Adaptive LLM  & all surfaces, 10 sessions  & 10 & 0 & 0\% \\
    \bottomrule
  \end{tabular}
\end{table}

The harness includes data-integrity safeguards so that every reported row is
independently reproducible. Each cell runs against an isolated
copy of the Maestro sweep directory, so a model's tuning-curve readback
reflects only its own simulation and cannot be contaminated by a neighbouring
cell. Subprocess result capture is locale-robust, enforcing UTF-8 end to end so
that remote simulation output is read back correctly regardless of host
encoding. Together these guarantee that every value in
Tables~\ref{tab:lcvco}--\ref{tab:failcounts} and Figure~\ref{fig:kvco} is
attributable to a single, re-runnable cell.

Because the contribution is a safety-bounded engineering system rather than a
single algorithm, each claimed guarantee maps to a concrete, inspectable
artifact in the released package (Table~\ref{tab:evidence}).

\begin{table}[t]
  \centering
  \caption{Evidence trace: each framework guarantee maps to a concrete artifact
  in the reproducibility package.}
  \label{tab:evidence}
  \small
  \begin{tabular}{@{}ll@{}}
    \toprule
    Guarantee & Evidence artifact \\
    \midrule
    Strict JSON contract & contract-repair / violation unit tests \\
    Maestro writeback    & \metric{writeback\_status}=\metric{ok} in matched runs \\
    Sweep isolation      & per-cell sweep-dir copy; no duplicated model rows \\
    Leak gate            & P0 gate pass: no path, PDK token, or API key \\
    Active leak resistance & 0\% ASR over fixed/live/adaptive attacks (Table~\ref{tab:redteam}) \\
    OP readback          & whitelisted operating-point scalar table, scrubbed \\
    \bottomrule
  \end{tabular}
\end{table}

\section{Discussion}
\label{sec:discussion}

\framework{} deliberately optimizes a conservative surface: sizing variables,
analysis setup, outputs, and specs. It is not a topology-synthesis system, and
it does not claim that cloud LLMs should see raw netlists or model cards. This
limited scope is a strength for industrial adoption because it matches a
common designer workflow: keep the schematic and PDK local, let an assistant
propose numeric sizing moves, and use the simulator as the source of truth.

The main limitation is statistical breadth. The matched comparison is one
audited run per model per task, and the evidence runs already show that both
budget and endpoint state can flip individual outcomes
(Sections~\ref{sec:ablation} and~\ref{sec:failures}); multi-seed replication and at least one
additional topology class are the clear next step, and the protocol---
documented reset, fixed Maestro tests, logged terminal causes---was designed
so that such replication is mechanical. The second limitation is that safety
and usefulness trade off against each other. Too little feedback leaves the
model guessing; too much feedback risks leaking design or PDK information.
The current evidence suggests that sanitized topology intent plus numeric
operating-point scalars plus curve-level summaries is a useful middle ground,
but channel-by-channel ablations across models remain to be quantified.
Third, the threat model bounds what we claim: provider-side retention of
first-emission reasoning traces, timing side channels, aggregate parameter
inference from operating-point scalars, and re-identification of a design from
its anonymized topology graph are explicitly out of scope for a client-side
framework (Table~\ref{tab:ladder}). The boundary constrains what each response
contains, not what a provider can infer by correlating many responses over a
run; closing that gap is what on-premise inference, and not a client-side
scrubber, is for.

Finally, model comparisons should not be interpreted as permanent rankings.
Provider endpoints, context behavior, tool-call discipline, and pricing change
over time---our own data contains one checkpoint whose outcome differs across
five days for endpoint reasons alone. The durable contribution is the
benchmark method: fixed reset, fixed Maestro setup, fixed spec, fixed feedback
schema, and logged failure reason. The reported tables are a snapshot
generated under that method, and the method is what we expect to outlive any
particular row.

\section{Conclusion}
\label{sec:conclusion}

This paper presented \framework, an NDA-safe closed-loop framework for
LLM-driven analog circuit optimization through industrial Virtuoso/Maestro flows.
The framework combines an explicit threat model with audited enforcement
status, scoped SKILL access through 28 whitelisted entry points,
PDK/path/model scrubbing on every return path, Maestro setup recipes with
landing verification, AC/DC and swept-metric readback, whitelisted
operating-point summaries, a strict JSON contract with six stop conditions,
and best-so-far writeback. In matched eleven-model benchmarks from a
documented reset state, both tasks run under the same three-corner
process--voltage--temperature sign-off with worst-corner gating: 7 of 11 models
pass the LC-VCO tuning-curve task, and 4 of 11 pass the harder op-amp task,
whose phase-margin gate rejects high-gain but unstable designs. Feedback-path
ablations show that individual sanitized channels are load-bearing, and the
failure taxonomy shows that the dominant failure is partial optimization of
visible metrics (worst-corner frequency placement for the LC-VCO, coupled gain
and stability for the op-amp) rather than protocol collapse. These results support the paper's central
claim: cloud LLMs can be evaluated and used in real analog EDA loops without
exposing raw PDK content, but their reliability must be measured at the level
of complete tool-loop behavior---contract discipline, endpoint stability, and
metric-driven repair---rather than standalone text reasoning.

\bibliographystyle{ACM-Reference-Format}
\bibliography{refs}

\end{document}